\documentclass[prb,twocolumn,showpacs,preprintnumbers,amsmath,amssymb]{revtex4}

\usepackage{graphicx}
\usepackage{dcolumn}
\usepackage{bm}

\begin{document}

\preprint{}

\title{Field-induced magnetic ordering in the Haldane system PbNi$_2$V$_2$O$_8$
}

\author{N. Tsujii, O. Suzuki, H. Suzuki, H. Kitazawa, G. Kido}
\affiliation{
National Institute for Materials Science, Sengen 1-2-1, Tsukuba 305-0047, Japan}%


\date{\today}

\begin{abstract}
The Haldane system PbNi$_2$V$_2$O$_8$ was investigated
by the temperature dependent magnetization $M(T)$ measurements
at fields higher than $H_{\rm c}$, with $H_{\rm c}$ the critical fields
necessary to close the Haldane gap.
It is revealed that $M(T)$ for $H > H_{\rm c}$ exhibits a cusp-like
minimum at $T_{\rm min}$, below which $M(T)$ increases with decreasing $T$
having a convex curve.
These features have been observed for both $H \parallel c$ and $H \perp c$,
with $c$ axis being parallel to the chain.
These data indicate the occurrence of field-induced magnetic ordering around $T_{\rm min}$.
Phase boundaries for $H \parallel c$ and $H \perp c$ do not
cross each other, consistent with the theoretical calculation for 
negative single-ion anisotropy $D$.
\end{abstract}

\maketitle

\section{Introduction}
Quantum spin systems with energy gaps have attracted considerable attention
because of their rich variety of interesting phenomena.
In these systems, the ground state is a nonmagnetic singlet state,
and there exists a finite energy-gap between the ground and the excited magnetic states.
Under certain magnetic fields $H_{\rm c}$, the energy of one of the magnetic excited states
becomes lower than that of the singlet state,
where nonmagnetic-magnetic crossover occurs.
Recently, special attention is paid to the magnetic transition
that develops just above $H_{\rm c}$ in these spin-gap systems.~\cite{Rice}
A number of novel phenomena is reported in these fields;
for example, magnetic ordering of TlCuCl$_3$ in fields is interpreted as
the Bose-Einstein condensation (BEC) of magnons.~\cite{Nikuni,Oosawa}
For the two dimensional dimer system SrCu$_2$(BO$_3$)$_2$,
superlattice formation of localized triplet is observed.~\cite{Kodama}

The Haldane systems, i.e., quasi-one-dimensional Heisenberg antiferromagnets with integer spins,
also are ones of the most  extensively studied spin gap systems.
For this class, unfortunately, field-induced magnetic ordering 
has hardly been observed experimentally.
The archetypal Haldane-system, Ni(C$_2$H$_8$N$_2$)$_2$NO$_2$(ClO$_4$) (NENP),~\cite{Renard}
shows no evidence of field-induced ordering down to 0.2 K in fields up to 13 T.~\cite{Kobayashi92}
Instead, the existence of an energy gap was revealed even at $H_{\rm c}$.~\cite{Kobayashi92}
This is explained by the existence of a staggered field on Ni sites,
which arises because the principal axis of the $g$ tensor in NENP tilts
alternately.~\cite{Chiba,Fujiwara}
This fact results in the slow crossover from nonmagnetic to magnetically polarized state in NENP,
preventing from phase transition induced by fields.

So far, field-induced ordering in Haldane-systems was
reported only for two cases: in Ni(C$_5$H$_{14}$N$_2$)$_2$N$_3$(PF$_6$)
and Ni(C$_5$H$_{14}$N$_2$)$_2$N$_3$(ClO$_4$), abbreviated NDMAP and NDMAZ, respectively.
Field-induced transitions in these systems were demonstrated 
by specific heat~\cite{Honda97,Honda98,Honda01,Kobayashi01}
and neutron diffraction experiments.~\cite{Chen,Zheludev01E}
In the ordered state, interestingly, unusual spin excitations are observed.
ESR~\cite{Hagiwara03} and inelastic neutron-scattering experiments~\cite{Zheludev03} on NDMAP
have revealed the existence of three distinct excitations in the ordered phase.
This feature is quite different from those in a conventional Neel state,
where dominant excitations are the spin-wave modes.~\cite{Zheludev02A}
Field-induced ordered phase in Haldane-systems is thus expected to illustrate
much kind of novel physics, if much more examples are available.

Here the compound PbNi$_2$V$_2$O$_8$ would be another candidate for the Haldane
system where field-induced ordering can be observed experimentally.
PbNi$_2$V$_2$O$_8$ has a tetragonal crystal structure with Ni$^{2+}$ ($S=1$) ions 
forming a chain along the $c$-axis.
Magnetic susceptibility, high-field magnetization, and inelastic neutron scattering
experiments were performed and their results consistently suggest that this system
is a Haldane-gap system.~\cite{Uchiyama}
The spin-gap closes at $H^{\parallel}_{\rm c} = 14$ T and $H^{\perp}_{\rm c} = 19$ T,~\cite{Uchiyama}
where $H^{\parallel}_{\rm c}$ and $H^{\perp}_{\rm c}$ are the critical fields
applied parallel and perpendicular to the chain ($c$-axis), respectively.
These values of $H_{\rm c}$ are within experimentally accessible range.
Moreover, PbNi$_2$V$_2$O$_8$ is reported to exhibit
impurity-induced magnetic transition around 3 K.~\cite{Uchiyama}
This transition was found to be a long-range magnetic ordering 
by the neutron diffraction~\cite{Lappas}
as well as the specific heat measurements~\cite{Masuda}.
These facts suggest the relatively large interchain coupling, $J_1$.
In fact, the $D-J_1$ plot~\cite{Sakai90} (Sakai-Takahashi diagram) for this compound,
where $D$ is the single-ion anisotropy, 
suggests that PbNi$_2$V$_2$O$_8$ is in the spin-liquid (disordered) regime
but very close to the long-range ordered regime.~\cite{Zheludev00}  
Hence, one can expect that applying fields beyond $H_{\rm c}$
will result in the magnetic ordering.
In the present paper, we have investigated magnetic properties of PbNi$_2$V$_2$O$_8$
at $H > H_{\rm c}$ using temperature dependent measurements of magnetization in static fields
up to 30 T,
and have observed indication of magnetic ordering above $H_{\rm c}$.

\section{Experimental}
Field-oriented powder sample of PbNi$_2$V$_2$O$_8$ was prepared 
as in the first report,~\cite{Uchiyama}
since single crystalline samples are not yet available.
Powder sample of PbNi$_2$V$_2$O$_8$ was synthesized by a solid state reaction
from PbO (99.999\% pure), NiO (99.99\%) and V$_2$O$_5$ (99.99\%).
They were mixed and heated in air, 
firstly at 600$^\circ$C and subsequently at 750$^\circ$C for several days 
with intermittent grindings. 

Powder X-ray diffraction (XRD) pattern agrees well with the calculated pattern
based on the structure refined by the neutron diffraction experiments,~\cite{Lappas,Comment}
and no second phase was detected.
The powder was aligned by a magnetic fields (6 T) in stycast.
The orientation was checked by the (004) XRD peak.
The result confirmed that the $c$-axis aligns parallel to the magnetic fields,
as is reported previously.~\cite{Uchiyama}
In the following, we refer to magnetization measured under fields
parallel to the $c$-axis, as $M^{\parallel}$, and
to that under fields perpendicular to the $c$-axis, as $M^{\perp}$.

Magnetization was measured by an extraction method.
Magnetic fields up to 15 T were generated by a superconducting magnet.
Fields higher than 15 T were generated by a hybrid magnet at the Tsukuba
Magnet Laboratory.
For the measurements of magnetization as a function of magnetic fields,
the fields were swept at the rate about 0.3 Tesla per minute at the temperature of 1.5 K.
For the temperature-dependent measurements, magnetization was measured under
constant magnetic fields.

\section{Results and discussion}
\begin{figure}[t]
\begin{center}
\includegraphics[width=8cm]{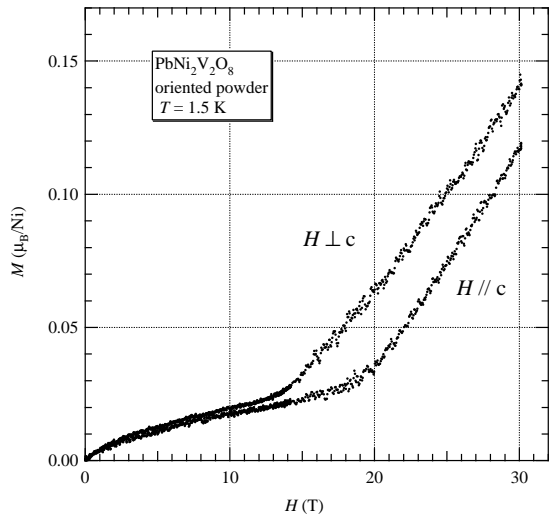}
\caption{\label{fig:epsart} Field dependence of the magnetization of the
PbNi$_2$V$_2$O$_8$ powder samples.}
\end{center}
\end{figure}
\begin{figure*}[t]
\begin{center}
\includegraphics[width=14cm]{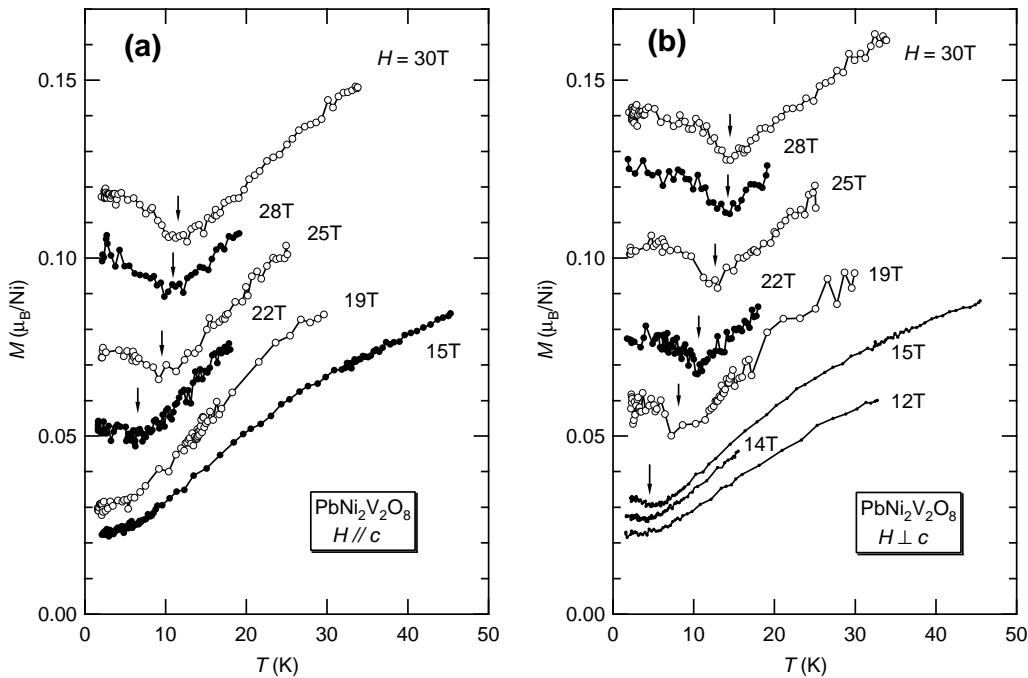}
\caption{\label{fig:epsart} Temperature dependence of the magnetization of 
PbNi$_2$V$_2$O$_8$ for (a) $H \parallel c$ and (b) $H \perp c$.
Arrows indicate $T_{\rm min}$ shown in the text.
}
\end{center}
\end{figure*}
Fig.1 shows field dependence of the magnetization $M^{\parallel}(H)$
and $M^{\perp}(H)$ measured at $T$ = 1.5 K.
Both the $M^{\parallel}(H)$ and $M^{\perp}(H)$ curves steeply increase above 
the critical fields, $H_{\rm c}^{\parallel}$ = 19 T and $H_{\rm c}^{\perp}$ = 13.5 T, respectively. 
These values correspond to the critical fields at which the Haldane gap closes,
and are in good agreement with the previous report obtained by
the pulsed-field experiments.~\cite{Uchiyama}

Here the $M(H)$ just above $H_c$ increases almost linearly with $H$.
This behavior differs from the theoretical predictions,
where $M(H)$ varies as proportional to $\sqrt{H-H_{\rm c}}$ for axially symmetric fields
($H \parallel c$).~\cite{Affleck,Takahashi}
One of the reason of this discrepancy can be the finite temperature effect.
The $\sqrt{H-H_{\rm c}}$ dependence easily becomes invisible by a slight thermal excitation,
as is observed in NDMAZ.~\cite{Kobayashi01}
Other origins can be imperfect powder orientation,
that can lead axially-asymmetric fields even for $H \parallel c$.
However, it is also possible that the linear $M(H)$ is intrinsic behavior of an
antiferromagnetically ordered system.
When the measured temperature is sufficiently low compared to the energy gap,
the system promptly enters the antiferromagnetic ordered regime above $H_c$,
as is shown below.

Fig.2(a) shows temperature dependence of the magnetization measured under
fields parallel to the $c$-axis, $M^{\parallel}(T)$.
For $H < H^{\parallel}_{\rm c} = 19$ T, the $M^{\parallel}(T)$ curves show no anomalies.
Below 5 K, $M^{\parallel}(T)$ have small but finite values (0.02-0.03$\mu_{\rm B}$),
and these are attributed to the saturation magnetization of impurity and/or defects.
For $H$ = 22 T, there exists a cusp-like minimum at
around $T_{\rm min}$ = 6.4 K. With increasing fields, $T_{\rm min}$ 
shifts to higher temperatures systematically.
$T_{\rm min}$ reaches to 11.5 K at $H$ = 30 T, as is seen in the figure.

Fig.2(b) shows temperature dependence of the magnetization 
measured under fields perpendicular to the $c$-axis, $M^\perp(T)$.
Similar to the above results, $M^{\perp}(T)$ also exhibits a cusp-like minimum
for $H > H^{\perp}_{\rm c}$ = 13.5T.

In both Fig.2 (a) and (b), $M(T)$ shows a sharp change in its slope at $T_{\rm min}$.
Moreover, below $T_{\rm min}$, $M(T)$ increases with decreasing $T$ showing a convex curve,
as is most clearly seen in $M^{\perp}(T)$ at $H$ = 30 T (Fig.2 (b)).
Such $M(T)$ curves quite resemble those field-induced magnetic ordering 
of the coupled dimers TlCuCl$_3$.~\cite{Oosawa,Nikuni}
In this compound, it is demonstrated that $T_{\rm min}$ at which $M(T)$ has a minimum
is the N\'{e}el temperature,
from the neutron diffraction~\cite{Tanaka} and the specific heat measurements.~\cite{Oosawa01}
Similarly, such cusp-like anomalies in $M(T)$ curves are shown to be 
the ordering temperature 
in the $S=1/2$ alternating chain Pb$_2$V$_3$O$_9$ (ref.~\cite{Waki})
and the quasi-2-dimensional BaCuSi$_2$O$_6$ (ref.~\cite{Jaime}) 
from specific heat measurements.
We hence conclude that the data shown in Fig.2 also demonstrate the occurrence of field-induced
magnetic ordering with N\'{e}el temperatures around $T_{\rm min}$.

It is notable that the Haldane system NDMAP exhibits a minimum in $M(T)$ for $H > H_{\rm c}$
at temperatures much higher than $T_{\rm N}$.~\cite{Honda01,HondaJAP}
The origin of the minimum is not yet clear, and possibly related to a
crossover into the low-temperature Tomonaga-Luttinger(TL) liquid regime,
as is predicted for non-interacting one-dimensional ladders.~\cite{Wang,Wessel}
This is purely a one-dimensional phenomenon,
and the three-dimensional ordering occurs at much lower temperatures.~\cite{Wessel}
For those cases, $M(T)$ curves around $T_{\rm min}$ are characterized by the relatively
broad minimum and the concave curve.~\cite{Honda01,HondaJAP,Wessel}
This is in clear contrast with the cusp-like anomalies and the convex curve below $T_{\rm min}$
in the present study as well as in those reported for TlCuCl$_3$ etc.,
which signals the 3-dimensional magnetic ordering.
It is of course important to perform other experiments in order 
to verify the magnetic ordering at $T_{\rm min}$.
The lack of single crystalline samples makes it difficult
to measure the specific heat of this anisotropic compound.
We   are then planning to measure the NMR spectra at high fields.

It may be interesting to compare the ordered-state induced by fields in PbNi$_2$V$_2$O$_8$
with that induced by impurity-doping in PbNi$_{2-x}$Mg$_x$V$_2$O$_8$.~\cite{Uchiyama}
It is shown that the ordered-state of the latter has inhomogeneous distribution of magnetic moment,
by ESR~\cite{Smirnov} and ${\rm \mu}$SR experiments.~\cite{Lappas}
In addition, the impurity-induced ordered state vanishes at fields higher than $H$ = 4 T,
where the Haldane state with an energy gap recovers.~\cite{Masuda}
In contrast, ordered state observed in the present experiments shows up only
above $H_{\rm c}$.
The largest value of $T_{\rm min}$ in the present study is $\sim$10 K for $H$ = 30 T,
which is much larger than the maximum value of $T_{\rm N}$ induced by Mg-doping,
3.3 K,~\cite{Smirnov} or the value of $zJ_1 \sim 0.03J \simeq 3.1 $K,
with $z$ the number of nearest chains, and $J$ the intrachain coupling.~\cite{Zheludev00}
This fact implies that the field-induced ordering occurs via the
developed antiferromagnetic-correlation along the chain,
and the ordered moment induced by fields is distributed
uniformly on the chain.

In Fig.3, the values of $T_{\rm min}$ are plotted against the applied fields.
This corresponds to the magnetic phase diagram for PbNi$_2$V$_2$O$_8$.
For both of $H \parallel c$ and $H \perp c$, $T_{\rm min}$ increases with fields.
It is notable that the phase boundaries for $H \parallel c$ and $H \perp c$ 
do not cross each other at least within the field range measured.
This is in qualitative agreement with the theoretical calculation
by Sakai,~\cite{Sakai01} the $HT$ phase diagram calculated for a Haldane chain with
negative $D$.
Indeed, $D/J = -0.05$ is estimated for PbNi$_2$V$_2$O$_8$ from inelastic
neutron scattering experiments.~\cite{Zheludev00}
In contrast, it is reported that crossing of the phase boundaries occurs in the
phase diagram of NDMAP and NDMAZ,~\cite{Honda98,Kobayashi01}
and is well explained by the theoretical calculation for positive $D$.~\cite{Sakai00}
Thus, PbNi$_2$V$_2$O$_8$ is the first example of field-induced order in 
the Haldane system with negative $D$.

\begin{figure}[tp]
\begin{center}
\includegraphics[width=8cm]{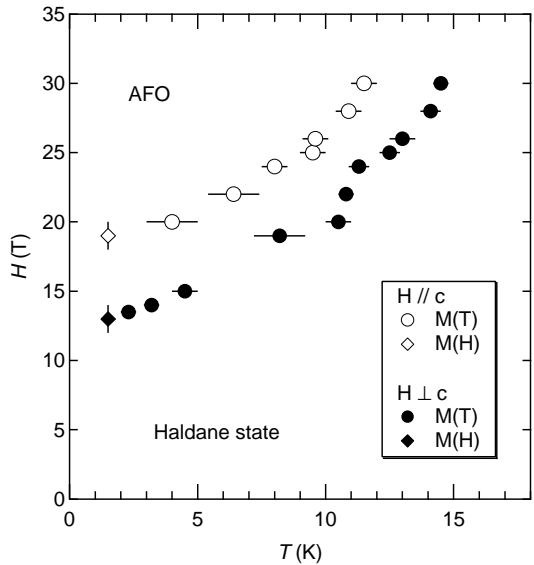}
\caption{\label{fig:epsart} Magnetic phase diagram of PbNi$_2$V$_2$O$_8$
suggested from the present experiments (symbols).
`AFO' and `Haldane-state' represent the antiferromagnetically ordered state
and the nonmagnetic spin-singlet state with a Haldane-gap, respectively.
Open symbols represent the data for $H \parallel c$, and filled ones are 
for $H \perp c$. Circles indicate $T_{\rm min}$ determined from the $M(T)$ curves,
while diamonds indicate $H_{\rm c}$ estimated from the $M(H)$ curves.
 }
\end{center}
\end{figure}
Since the early stage of the research of the Haldane systems,
its magnetic state at $H > H_{\rm c}$ has been discussed theoretically
in terms of the BEC picture.~\cite{Affleck,Sorensen}
In the following, we discuss on the possible condensed state in the ordered phase.
In Fig.2 (a), one can see that $M^{\parallel}(T)$ increases 
below $T_{\rm min}$ with decreasing $T$.
Such increase cannot be explained by a conventional mean-field theory,~\cite{Tachiki}
which predicts almost flat $M(T)$ below the ordered temperature.
Instead, this increase is successfully explained by the magnon BEC theory,
as to be due to the increase of magnon number as the condensation sets in.~\cite{Nikuni}
To apply the magnon BEC theory to our case, it is essential that the rotational
symmetry around the magnetic field be conserved.~\cite{Nikuni}
In the present case, $M^{\parallel}(T)$ holds this restriction.

It is rather surprising that the $M^{\perp}(T)$ also increases below $T_{\rm min}$
as is seen in Fig.2(b).
Here $H$ is applied perpendicularly to $D$,
thereby the rotational symmetry of the Hamiltonian around $H$ is broken.
In such cases, magnon BEC picture is not assured because the number of bosons 
is not conserved.~\cite{Nikuni}
In fact, the $M^{\perp}(T)$ of NDMAP does not increase below $T_{\rm N}$
but becomes flat against $T$.~\cite{Honda01,HondaJAP}
Such behavior is consistent with the Ising-like antiferromagnet that
is predicted to develop for $H \perp D$.~\cite{Sakai00}
For the present system, the similarity of $M^{\perp}(T)$ and $M^{\parallel}(T)$
may be due to the relatively small $D$ ($D/J = -0.05$).
This point should be studied more carefully.

It should be remarked, however, that the BEC picture requires some rigorous conditions.
First, the concentration of magnons must be enough dilute.~\cite{Nikuni}
In fact, experiments on KCuCl$_3$, isostructural of TlCuCl$_3$,
showed that the $M(T)$ curve becomes flat below $T_{\rm N}$ for 
fields well above $H_{\rm c}$.~\cite{Oosawa02}
This behavior implies that for this dense magnon condition, 
mean field approximation is a better description.
The BEC picture should hence be applied only for the region $H - H_{\rm C} \sim 0$.
Moreover, it is recently argued that some anisotropic interactions
arising from spin-orbit coupling like the Dzyaloshinsky-Moriya(DM) interaction 
and/or the staggered $g$ effect can be qualitatively modify the BEC description,
even if the interactions are very weak.~\cite{Sirker04,Sirker05}
Recent ESR measurements on TlCuCl$_3$ have indeed suggested the existence of 
such interactions.~\cite{Glazkov}
For the present case, the screw-like crystal structure of PbNi$_2$V$_2$O$_8$
may cause the DM interaction, as is suggested to explain the weak-ferromagnetism 
in the isostructural SrNi$_2$V$_2$O$_8$.~\cite{Zheludev00}

\section{Conclusions}
We have observed cusp-like anomaly at $T_{\rm min}$ in the $M(T)$ curves for $H > H_{\rm c}$.
The value of $T_{\rm min}$ increases with applied fields.
These observations suggest the evolution of field-induced magnetic ordering
in the Haldane chain system PbNi$_2$V$_2$O$_8$.
Magnetic phase diagram of this system up to 30 T is presented.
The phase boundaries for $H \parallel c$ and $H \perp c$ do not
cross each other, in qualitative agreement with the $HT$ phase diagram calculated theoretically for a Haldane 
system with $D<0$.~\cite{Sakai01}

In the ordered phase, it is revealed that the magnetizations increase with decreasing $T$
and the $M(T)$ have a convex curve for both the directions $H \parallel c$ and $H \perp c$.
These features may support that the magnon Bose-Einstein condensation picture
can be applicable as an approximation for Haldane gap systems at least for $H \parallel c$.
However, possible anisotropic effects including the Dzyaloshinsky-Moriya interaction
can modify the description of the ordered state significantly.

\begin{acknowledgments}
N.T. gratefully acknowledges M. Hagiwara, A. Oosawa, M. Hase, and H. Kageyama for fruitful discussions.
He also thanks T. Waki and K. Yoshimura for informing their results, and 
K. Hashi and H. Shinagawa for the help of field-oriented sample preparation.

\end{acknowledgments}


\begin{thebibliography}{00}
\bibitem{Comment} The powder x-ray diffraction pattern was calculated based on the
parameters listed in ref.~\cite{Lappas}, but the occupancy of 1.0 for the 8$a$ site (Pb site)
was employed.
\end{thebibliography}
\end{document}